\documentclass{epl}

\title{Bubble kinetics in a steady-state column of aqueous foam}
\shorttitle{Bubbles in steady foam}
\author{K. Feitosa, Olivia L. Halt, Randall D. Kamien, \and D. J. Durian}

\institute{
  Department of Physics and Astronomy, University of Pennsylvania,
  209 S. 33$rd$ Street, Philadelphia, PA 19104-6396, USA\\
}

\pacs{83.80.Iz}{Emulsions and foams}
\pacs{82.70.Rr}{Aerosols and foams}

\begin{document}

\maketitle

\begin{abstract}
We measure the liquid content, the bubble speeds, and the
distribution of bubble sizes, in a vertical column of aqueous foam
maintained in steady-state by continuous bubbling of gas into a
surfactant solution.  Nearly round bubbles accumulate at the
solution/foam interface, and subsequently rise with constant speed.
Upon moving up the column, they become larger due to gas diffusion
and more polyhedral due to drainage. The size distribution is
monodisperse near the bottom and polydisperse near the top, but
there is an unexpected range of intermediate heights where it is
bidisperse with small bubbles decorating the junctions between
larger bubbles. We explain the evolution in both bidisperse and
polydisperse regimes, using Laplace pressure differences and taking
the liquid fraction profile as a given.
\end{abstract}

Aqueous foam is a quintessential non-equilibrium system, even in
absence of film rupture.  An initially homogeneous foam will drain
due to gravity, and will coarsen due to gas diffusion, en route to
an equilibrium state of total phase separation where the foam
vanishes. The beautiful topology and microstructure of soap films
and their junctions into Plateau borders and vertices have inspired
wide-ranging studies of coarsening and drainage as fundamental
evolution mechanisms \cite{Stavans93, WeaireBook99}. However, drier
foams coarsen more rapidly, and coarser foams drain more rapidly,
and this interplay enhances temporal evolution and spatial
inhomogeneity in situations such as free \cite{ArnaudEiffel00} and
forced drainage \cite{HutzlerCoarsening00}. In spite of good
progress, several open questions remain.  For example there is no
consensus on how the coarsening rate depends on liquid content
\cite{HutzlerCoarsening00,Hilgenfeldt01,MoinPRL02}. And there is
little understanding of how drainage is affected by a distribution
of bubble sizes.

To advance the understanding of such issues we examine a geometry in
which a vertical column of foam is created by a continuous stream of
small bubbles into a pool of surfactant solution. Here the foam
reaches a state where the bubbles rise at constant speed, and the
liquid remains at rest, in the laboratory frame. Prior studies of
such steady-state foams focus mainly on the height of the foam as
the key observable quantity \cite{Pilon01, Barbian03}, though
recently the liquid fraction profile was predicted under the
assumption of constant bubble size \cite{Grassia06}. Here the liquid
content, the bubble speed, and the bubble size distribution, are all
independent of time and are hence measured at our leisure as a
function of height. This is a simplification over free and forced
drainage, which require study as a function of both position and
time. Furthermore, the steady-state condition allows for a different
and unexplored type of interplay between drainage and coarsening. As
presented below, our measurements reveal two striking qualitative
features that can be quantitatively modeled: the formation of a
bidisperse bubble size distribution, and a rate of coarsening that
increases without apparent bound for decreasing liquid fractions.

\section{Methods and data}

We generate a steady-state column of foam by continuously blowing
gas into a surfactant solution inside a tall Lucite cylinder, 61~cm
in height and 5.08~cm in inner diameter. The bottom of the cylinder
is filled with 120~ml of an aqueous solution of AOS ($\alpha$-olefin
sulfonate, Bio-Terge AS-40 CG-P, Stepan Company) and NaCl with
concentrations of 0.4\% and 0.01\% by weight respectively.  The
solution has surface tension $\gamma=44$~dyne/cm and viscosity
$\mu=0.011$~g/cm-s. Small bubbles of $\mathrm{CO}_2$ (Airgas East,
99.2\% pure, H$_2$O$<200$~ppm, NH$_3<25$~ppm, CO$<10$~ppm,
H$_2$S/SO$_2<5$~ppm, NO$_X<2.5$~ppm) are sparged into the solution
through a fine mesh cloth. This gas has solubility 1.17~mol/l and
diffusivity $1.8\times10^5$~cm$^2$/s in water. The flow rate of
$\mathrm{CO}_2$ is held fixed and is measured by an electronic flow
meter (Omega, model FMA 3102) to be $0.505$~ml/s; the resulting gas
flux is $0.029\pm0.001$~cm/s, where the error is dominated by
calibration uncertainty. After six hours, a steady state is achieved
in which the foam is about 45~cm high and where the input of small
bubbles at the bottom is balanced by the bursting of large bubbles
into the atmosphere at the top. Visually, there is no bursting
anywhere in the foam except at the top free surface.

\begin{figure}
\onefigure[width=0.8\textwidth]{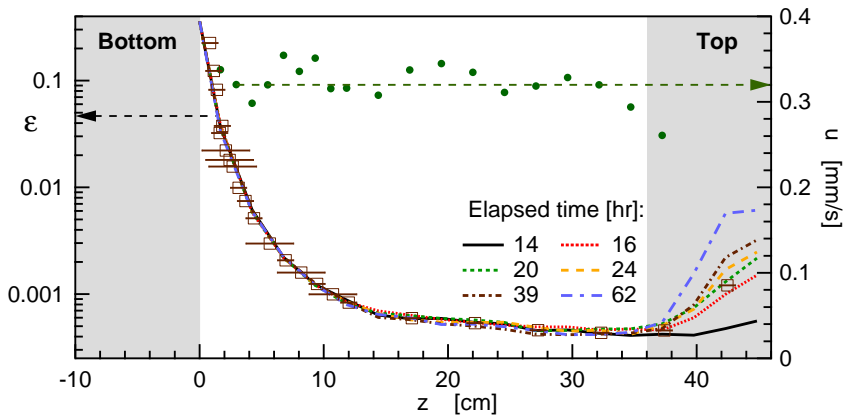} \caption{Liquid fraction
$\varepsilon$ (left) and average bubble speed $u$ (right) as a
function of height. The liquid fraction data are for two runs; the
first is vs time (curves), the second is for varied electrode
heights (squares); horizontal error bars indicate the range over
which conductivity was measured. The bottom gray area represents the
surfactant solution; the top gray area represents the region where
bubbles burst.} \label{euz}
\end{figure}

The physical properties of the steady-state foam are measured as
follows.  First, the liquid volume fraction $\varepsilon(z)$ is
deduced from electrical conductivity vs height $z$. A series of 19
stainless steel electrodes, 1.27~cm in diameter and 2.54~cm apart,
are attached to two acetal strips placed opposite one another inside
the foam column. The conductivities of liquid and foam are both
measured by an impedance meter (1715 LCR Digibridge, QuadTech)
configured to measure the resistance of a parallel
resistor-capacitor equivalent circuit, operated at a frequency of
1~kHz and voltage level of~1.00 V. At this frequency, the capacitive
contribution of the foam is negligible.  The conductivity is
measured at smaller intervals than 2.54~cm, both by connecting
diagonally-opposite electrodes and by changing the height of the
pool of surfactant solution relative to the electrode heights.  The
liquid fraction is calculated from the ratio of foam to liquid
conductivity, $\sigma=\sigma_{foam}/\sigma_{liquid}$, using the
semi-empirical relation $\varepsilon=3\sigma(1 + 11\sigma)/(1 +
25\sigma + 10\sigma^2)$ \cite{FeitosaJPCM05}. Results for
$\varepsilon(z)$ are displayed on the left-hand axis in
Fig.~\ref{euz}, for several different ages of the sample.  Indeed
the data are independent of time, indicative of a steady-state.  The
foam is wetter on the bottom and drier towards the top.  At the very
top, where the bubbles burst, surfactant gradually accumulates and
causes an uncontrolled increase in liquid conductivity; this gives
an erroneous apparent increase in liquid fraction with time.

\begin{figure}
\onefigure[width=0.8\textwidth]{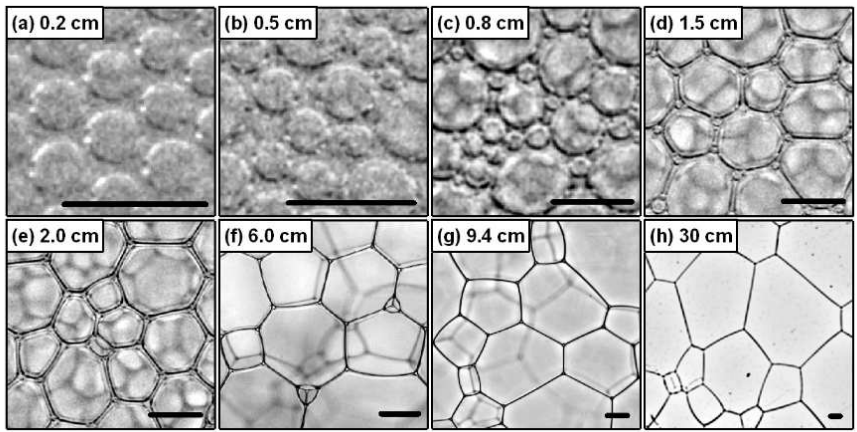} \caption{Example surface
images at different heights, as labeled; scale bars are 1~mm. These
images were acquired after steady-state was established.}
\label{bubbles}
\end{figure}

Next, the upward speed $u(z)$ of the bubbles at a given height is
deduced from a rapid sequence of digital photographs (Nikon D70; AF
Micro Nikkor lens, 60 mm, 1:2.8D).  Example images are given in
Fig.~\ref{bubbles}.  The average bubble displacement is calculated
as the location of the peak of the vertical cross-correlation of two
subsequent images.  The time between images is 0.333~s, chosen so
that the bubbles move noticeably without significant change in the
field of view or the packing structure. Values for ten successive
pairs of images are averaged together. Final bubble speed data are
displayed vs height on the right-axis of Fig~\ref{euz}.  The speeds
are constant, $u=0.032\pm0.003$~cm/s, independent of height. This
value is consistent with the directly-measured gas flux,
$0.029\pm0.001$~cm/s. Furthermore, we observe no global convection
or swirling. Therefore, the bubble motion is a constant plug-flow
across the column.

\begin{figure}
\onefigure[width=0.8\textwidth]{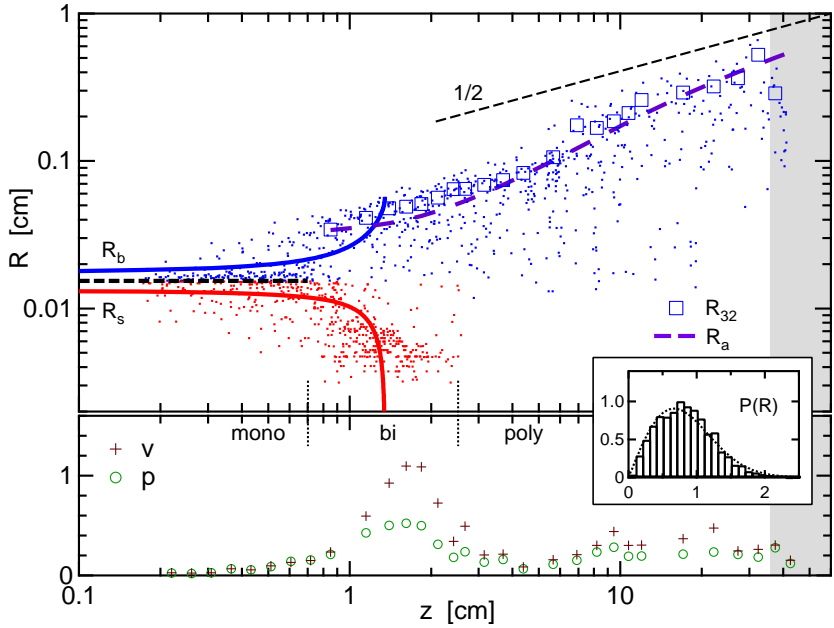} \caption{TOP: Scatter plot of
individual bubble radii vs height; ratios of third to second moments
of the distribution, $R_{32}$, are shown by open squares. The
average radius at the very bottom is $\langle R\rangle=0.0154$~cm,
as given by the horizontal dashed black line. The standard deviation
at the very bottom is $\sigma_o=0.0022$~cm. The solid blue and red
curves, which represent solution of
Eqs.~(\protect\ref{cons}-\protect\ref{bisoln}), approach $\langle R
\rangle \pm \sigma_o$ at $z=0$.  The dashed purple curve represents
solution of Eq.~(\protect\ref{Ra}). The gray area represents the top
of the sample, where bubbles burst. BOTTOM: Height dependence of the
polydispersity parameters $v=\langle R^2\rangle / \langle
R\rangle^2-1$ and $p=\langle R^3\rangle^{2/3} / \langle R^2 \rangle
- 1$.  INSET: Distribution of bubble radii in the polydisperse
region; the x-axis is scaled by $R_{32}$, which is related to the
average radius by $R_{32}/\langle R\rangle = 1.38\pm0.13$.  The
dashed curve is a fit to a Weibull distribution, $\propto R \exp(-k
R^2)$.} \label{radii}
\end{figure}

Finally, image analysis is performed to identify individual bubbles
and the area $A$ they occupy on the column surface. The equivalent
radius is taken as $R=\sqrt{A/\pi}$; results are not adjusted
according to Ref.~\cite{ChengIECF83}, which assumes that the surface
bubbles are equivalent to a random cut in a bulk foam. For heights
greater than $z=1$~cm (which includes both bidisperse and
polydisperse regimes - see below), we are able to inspect the region
2-3 bubbles from the surface; we notice no qualitative difference
from boundary bubbles.  A scatter plot of radius vs height is
displayed in Fig.~\ref{radii}, with one small point for each
measured bubble. For heights $0<z<0.7$~cm near the bottom of the
column the bubbles are small and the distribution is relatively
monodisperse; the average and standard deviation are $\langle
R\rangle=0.0154$~cm and $\sigma_o=0.0022$~cm. For increasing heights
$z>2.5$~cm through most of the column, the bubbles become
progressively larger and the size distribution broadens.  The
breadth of the distribution can be gauged by the dimensionless
variance $v=\langle R^2\rangle / \langle R\rangle^2-1$, and by the
polydispersity parameter $p=\langle R^3\rangle^{2/3} / \langle R^2
\rangle - 1$ \cite{KraynikPRL04}. These quantities are shown vs
height in the bottom plot of Fig.~\ref{radii}.  Both start small,
develop a transient overshoot, and become constant at large heights,
$v=0.27\pm0.11$ and $p=0.18\pm0.06$. For $z>3$~cm we therefore scale
the individual radii by a running average, and combine all the data
into a single dimensionless size distribution.  The result
(Fig.~\ref{radii} inset) is similar to Fig.~18 of
Ref.~\cite{GlazierGrestPMB00}, and is well-fit by a Weibull form.

For a narrow range of intermediate heights, roughly
0.7~cm$<z<2.5$~cm, the bubble radii data in Fig.~\ref{radii} fall
into two distinct branches. The size distribution doesn't smoothly
broaden as a function of height in going from monodisperse to
polydisperse regions; rather, it develops a bidisperse region and
becomes polydisperse only after the branch of small bubbles
disappears. Note that the existence of a bidisperse region can be
seen directly by inspection of surface images. At the bottom,
Figs.~\ref{bubbles}a-b, the bubbles are monodisperse; towards the
top, Figs.~\ref{bubbles}f-h, the bubbles are polydisperse with
larger and smaller bubbles arrayed at random; in between,
Figs.~\ref{bubbles}c-e, the junctions between big bubbles are all
decorated with small bubbles.  The small ``decorating'' bubbles
become progressively smaller in Figs.~\ref{bubbles}d-e, and may best
be viewed by enlarging the figure on-screen.

This surprising observation has no precedent, to our knowledge.
Prior experiments on the evolution of initially-monodisperse bulk
foams all found smooth relaxation to a polydisperse size
distribution with no intermediate bidisperse regime
\cite{HutzlerCoarsening00, Magrabi99, GananCalvoAPL04}. Perhaps the
closest connection is with Fig.~2 of Ref.~\cite{GlazierPRA87}, where
coarsening was depicted for an initially-ordered two-dimensional
foam. As here, that sample developed two distinct populations and
became uniformly polydisperse only after the population of small
bubbles fully disappeared. Furthermore the second moment $\mu_2$ of
the side distribution did not increase monotonically from a small
value, but rather developed a transient overshoot at intermediate
times \cite{StavansPRL89}, much like the polydispersity parameter
here.  However, by contrast with our sample, the small bubbles and
big bubbles were separately clustered together in segregated
domains.

\section{Models}

We now attempt to model the observed behavior of bubble size vs
height in the column, with the measured liquid-fraction profile as a
given. Since the vertical bubble speed is a constant equal to the
gas flux per unit area, $u=0.032$~cm/s, and since there is no
convection or film rupture, the basic physics is controlled solely
by the diffusion of gas from smaller to larger bubbles. Ultimately
this is driven by surface tension, where the smaller bubbles are
under higher Laplace pressure, and is a means of reducing total
interfacial area. Furthermore, time- and height- derivatives are
related by ${\rm d}/{\rm d}t=u{\rm d}/{\rm d}z$. The approach
developed by Lemlich \cite{LemlichIECF78} and by Markworth
\cite{Markworth85} is similar to earlier work on the kinetics of
phase separation in binary liquids and metal alloys. The rate of
change of a bubble's volume is proportional to its surface area and
to its Laplace pressure difference with respect to a certain mean or
critical bubble size: ${\rm d} R^d/{\rm d}t \propto R^{d-1} (1/R_c -
1/R)$, or ${\rm d}R/{\rm d}t \propto (1/R_c - 1/R)$, for any
dimension $d$; thus large bubbles $R>R_c$ grow while small bubbles
$R<R_c$ shrink. Note that this is a ``mean-field'' theory, in which
each bubble gets its marching orders by comparison with an average
rather than by comparison with its actual neighbors. The next step
is to combine this rate of change with an expression of total gas
conservation to arrive at a differential equation for the evolution
of the distribution of bubble sizes.  The result predicts that an
initially-narrow distribution will smoothly broaden until reaching a
stationary state, in which the shape of the distribution is constant
and the average bubble radius grows as the square-root of time. In
particular, note that this approach does not predict a bidisperse
regime at intermediate times.

In the stationary polydisperse regime, the evolution of the average
bubble radius may be modeled as ${\rm d}R_a/{\rm d}t=D_o
F(\varepsilon)/R_a$.  Here $D_o$ is a materials constant with units
of cm$^2$/s; it is proportional to the gas diffusivity and
solubility, and is inversely proportional to film thickness. Three
different forms have been suggested for the liquid-fraction
dependence of the coarsening rate:
$F(\varepsilon)=1-\sqrt{\varepsilon/0.36}$
\cite{HutzlerCoarsening00},
$F(\varepsilon)=(1-\sqrt{\varepsilon/0.44})^2$ \cite{Hilgenfeldt01},
and $F(\varepsilon)=1/\sqrt{\varepsilon}$ \cite{MoinPRL02}.  The
former two are based on total blockage of diffusion by the Plateau
borders; the latter is based on scaling of pressure differences with
the average Plateau border curvature.  Direct measurements of the
coarsening rate \cite{MoinPRL02} support
$F(\varepsilon)=1/\sqrt{\varepsilon}$ but do not definitively rule
out the other two forms.  However, our new scatter data for bubble
radius vs height contradict these with no analysis whatsoever. If
the coarsening rate were constant for very dry foams, as these
suggest, then we'd expect ${\rm d}R_a/{\rm d}t\propto 1/R_a$; this
implies $R_a(t)\sim t^{1/2}$ and hence $R_a(z)\sim z^{1/2}$ in our
experiment.  Since the bubble radii data in Fig.~\ref{radii} grow
faster than $z^{1/2}$, the coarsening rate must continue to increase
for ever-drier foams.

Altogether our observations in the polydisperse regime are thus best
modeled by
\begin{equation}
    R_a{\rm d}R_a/{\rm d}t = D_o/\sqrt{\varepsilon},
\label{Ra}
\end{equation}
where $D_o$ is the only unknown parameter.  It has been suggested
that the most suitable average radius is given by the ratio of third
to second moments of the bubble radius distribution, $R_{32}=\langle
R^3\rangle / \langle R^2 \rangle$, known as the Sauter mean radius
\cite{KraynikPRL04}. We compute this at each height for which the
liquid fraction was measured. Then we directly compute $D_o$ from
Eq.~(\ref{Ra}) as the height-average of $\sqrt{\varepsilon} R_{32}
(u {\rm d} R_{32}/{\rm d}z)$.  The result is $D_o=(3\pm1)\times
10^{-6}$~cm$^2$/s. For illustration, this value was used with
Eq.~(\ref{Ra}) to generate the dashed purple curve in
Fig.~\ref{radii} giving the growth of the Sauter radius starting
from the observed value at the smallest height measured. This
prediction grows faster than $z^{1/2}$ and agrees quite well with
$R_{32}$ data throughout the column.

Now we turn attention to the mono- and bidisperse regimes near the
bottom of the column.  The simplest model we can conceive that goes
beyond mean field is one in which there are precisely two sizes of
bubbles, such that each big bubble $R_b$ grow at the sole expense of
$N$ neighboring small bubbles $R_s$.  For a tractable model, we
imagine a one-dimensional situation in which gas conservation is
expressed as
\begin{equation}
    NR_s + R_b = (N+1)R_o,
\label{cons}
\end{equation}
and where the right-hand side remains constant as $R_s$ shrinks and
$R_b$ grows.  To complete the model, we take the time evolution of
the small bubbles according to the Laplace pressure difference and
the above liquid-fraction dependence as
\begin{equation}
    \frac{ \upd R_s}{ \upd t} = \frac{ \alpha D_o}{
    \sqrt{\varepsilon}} \left( \frac{1}{R_b} - \frac{1}{R_s}
    \right).
\label{bimodel}
\end{equation}
Here $\alpha$ is a dimensionless parameter whose value is expected
to be of order 1.  Substituting $R_b$ from Eq.~(\ref{cons}),
separating variables, and integrating with initial condition
$R_s(0)=R_o-\delta_o$, gives an exact solution for the evolution of
$R_s(t)$:
\begin{equation}
    \frac{\alpha D_o t}{\sqrt{\varepsilon}} = \frac{{R_o}^2}{N+1}
    \left\{
        \frac{(R_o-\delta_o-R_s)[N(R_o-\delta_o+R_s)-2R_o]}{2 {R_o}^2} +
        \ln\left(\frac{R_o-R_s}{\delta_o}\right)
    \right\}.
\label{bisoln}
\end{equation}
If $R_s$ is treated as the independent variable, then time $t$ and
height $z=ut$ can be generated from Eq.~(\ref{bisoln}) while $R_b$
can be generated from Eq.~(\ref{cons}).

We compare our model with data as follows, adjusting as few
parameters as possible.  The initial conditions are taken from the
average $\langle R\rangle=0.0154$~cm and standard deviation
$\sigma_o=0.0022$~cm of the bubble sizes at the very bottom of the
column, such that $R_s(0)=\langle R\rangle - \sigma_o=0.0132$~cm and
$R_b(0)=\langle R\rangle +\sigma_o=0.0176$~cm.  In light of gas
conservation Eq.~(\ref{cons}), this corresponds to $R_o=\langle
R\rangle - \sigma_o(N-1)/(N+1)$ and $\delta_o=2\sigma_o/(N+1)$.  The
value of $N$ is then deduced from a parametric plot of big vs small
bubble radii at equal heights in the bidisperse range.  The form of
$R_b$ vs $R_s$ is linear, and best fit gives $N=3.0\pm0.5$.
Altogether the parameters in the prediction Eq.~(\ref{bisoln}) are
thus fixed as $D_o=3\times10^{-6}$~cm$^2$/s, N=3, $R_o=0.0143$~cm,
and $\delta_o=0.0011$~cm.  The only unknown adjustable parameter is
$\alpha$.  The best fit gives $\alpha=0.4\pm0.1$, as illustrated by
the solid red and blue curves in Fig.~\ref{radii}.  The agreement
with the scatter data is quite satisfactory across the mono- and
bidisperse regimes.

\section{Conclusions}

In summary we observe the bubble size distribution in a steady-state
column of foam to evolve from monodisperse, to bidisperse, to
polydisperse, as a function of height.  The bidisperse region was
unexpected, but, once formed, its evolution may be understood in
term of the evaporation of small bubbles into neighboring large
bubbles. In addition we observe the rate of coarsening in the
polydisperse regime to increase with decreasing liquid fraction,
without detected bound, consistent with $R_a{\rm d}R_a/{\rm
d}t=D_o/\sqrt{\varepsilon}$. These findings expand the current
understanding of coarsening and shed light onto its interplay with
drainage phenomena.  Furthermore they underscore the importance of
the bubble size distribution, and the nontrivial role it plays. Our
findings also raise new questions. How does a bidisperse foam emerge
from a nearly monodisperse initial condition? It appears that the
fastest growing ``mode'' is one in which essentially alternate
bubbles shrink and grow in a spatially-correlated manner. Is this
generic, or is it due to extraordinary conspiracy of drainage and
coarsening in steady-state foams?  Could it be due to segregation of
less-soluble impurities in the gas \cite{WeairePageron}? Could it be
due to bidisperse pressures for equal-volume bubbles, as in the
Weaire-Phelan A15 foam \cite{WeairePhelanA15, LevineGrestPML96}?
This latter effect seems most likely to us, but suitably generalized
to random structures. Because of Plateau's rules all the films
cannot have a mean curvature of zero. For every bubble face which
bows out there is an inward face on the adjacent bubble. It follows
that there are two distributions of faces with negative and positive
curvature, respectively.  Thus in a dry foam with equal volume
bubbles, there must be a distribution of bubble curvatures and hence
pressures. This scenario emphasizes the lack of connection between
bubble size and pressure.  Resolving these issues would have
fundamental interest, and could aid applications in which
steady-state foams must be controlled for efficient fractionation of
chemicals and particulates.

\acknowledgments We thank A. Saint-Jalmes for helpful conversations.
This work was supported by NASA (NNC04GB61G, KF and DJD), NSF
(DMR05-47230, OLH and RDK), and the ACS Petroleum Research Fund (OLH
and RDK).

\end{document}